\RequirePackage{fix-cm} 
\documentclass[a4paper, twoside, reqno, dvips, 12pt]{amsart}
\usepackage{fixltx2e}   

\usepackage{etex}

\usepackage[latin1]{inputenc}
\usepackage[T1]{fontenc}

\usepackage{eucal}
\usepackage{esint}
\usepackage{dsfont}
\usepackage{xspace}
\usepackage{amsgen}
\usepackage{amsthm}
\usepackage{amssymb}
\usepackage{amsmath}
\usepackage{upgreek}
\usepackage{wasysym}
\usepackage{amsfonts}
\usepackage{stmaryrd}
\usepackage{mathtools}

\usepackage{mathrsfs}
\DeclareMathAlphabet{\mathscrbf}{OMS}{mdugm}{b}{n}

\usepackage{a4wide}

\usepackage[dvipsnames, table]{xcolor}
\definecolor{bckg}{RGB}{20.8, 20.8, 20.8}
\definecolor{oneblue}{rgb}{0.0, 0.0, 0.85}
\definecolor{Lightblue}{RGB}{214, 214, 214}
\definecolor{bluepigment}{rgb}{0.2, 0.2, 0.6}
\definecolor{charcoal}{rgb}{0.21, 0.27, 0.31}
\definecolor{denimblue}{rgb}{0.08, 0.38, 0.74}
\definecolor{Lightgray}{rgb}{0.89, 0.89, 0.89}
\definecolor{darkgrey}{rgb}{0.273, 0.281, 0.30}
\definecolor{darkelectricblue}{rgb}{0.33, 0.41, 0.47}

\usepackage{psfrag}
\usepackage{graphicx}
\usepackage{subfigure}
\usepackage{morefloats}
\usepackage{indentfirst}

\usepackage{acronym}
\usepackage{microtype}
\usepackage[labelsep=period,%
            labelfont={bf,sf,color=bluepigment},%
            justification=raggedright]{caption}

\usepackage[perpage, symbol]{footmisc}

\usepackage[usenames, dvipsnames]{pstricks}
\usepackage{epsfig}
\usepackage{pst-grad} 
\usepackage{pst-plot} 

\usepackage[colorlinks,
           urlcolor=oneblue,
           linkcolor=denimblue,
           citecolor=NavyBlue,
           bookmarksopen=false,
           pdfpagemode=UseNone,
           pagebackref]{hyperref}

\usepackage[sort&compress, comma, square, numbers]{natbib}

\usepackage[explicit]{titlesec}

\titleformat{\section}
  {\color{NavyBlue}\Large\sffamily\bfseries}
  {}
  {0em}
  {\colorbox{bckg!5}{\parbox{\dimexpr\linewidth-2\fboxsep\relax}{\centering\thesection. #1}}}
  [\vspace*{0.33em}]

\titleformat{name=\section,numberless}
  {\color{NavyBlue}\Large\sffamily\bfseries}
  {}
  {0.0em}
  {\colorbox{bckg!10}{\parbox{\dimexpr\linewidth-2\fboxsep\relax}{\centering#1}}}
  [\vspace*{0.33em}]

\titleformat{\subsection}
  {\color{NavyBlue}\large\sffamily\bfseries}
  {}
  {0.0em}
  {\colorbox{bckg!5}{\parbox{\dimexpr\linewidth-2\fboxsep\relax}{\centering\thesubsection. #1}}}
  [\vspace*{0.33em}]

\titleformat{name=\subsection,numberless}
  {\color{NavyBlue}\Large\sffamily\bfseries}
  {}
  {0em}
  {\colorbox{bckg!10}{\parbox{\dimexpr\linewidth-2\fboxsep\relax}{\centering#1}}}
  [\vspace*{0.33em}]

\titleformat{\subsubsection}
  {\color{bluepigment}\sffamily\normalsize\bfseries}
  {\thesubsubsection}
  {0.5em}
  {#1}
  [\vspace*{0.33em}]

\titleformat{\paragraph}[runin]
  {\color{bluepigment}\sffamily\small\bfseries}
  {}
  {0em}
  {#1}

\titlespacing{\section}{1.0em}{1.5em plus 2pt minus 2pt}%
{1.0em plus 2pt minus 2pt}[0em]
\titlespacing{\subsection}{1.0em}{1.5em plus 2pt minus 2pt}%
{1.0em}[0em]
\titlespacing{\subsubsection}{1.0em}{1.5em plus 2pt minus 2pt}%
{1.0em plus 2pt minus 2pt}[0em]

\usepackage{titletoc}

\contentsmargin{0.5em}
\newlength{\tocsep} 
\titlecontents{section}[\tocsep]
  {\addvspace{10pt}\bfseries\sffamily}
  {\contentslabel[\thecontentslabel]{\tocsep}}
  {}
  {\ \titlerule*[0.75pc]{.}\ \thecontentspage}
  []
\titlecontents{subsection}[\tocsep]
  {\addvspace{8pt}\sffamily}
  {\contentslabel[\thecontentslabel]{\tocsep}}
  {}
  {\ \titlerule*[0.5pc]{.}\ \thecontentspage}
  []
\titlecontents*{subsubsection}[\tocsep]
  {\addvspace{2pt}\footnotesize\sffamily}
  {}
  {}
  {\ \titlerule*[0.35pc]{.}\ \thecontentspage}
  [\\*]

\makeatletter
\def\@setauthors{%
  \begingroup
  \def\thanks{\protect\thanks@warning}%
  \trivlist
  \centering\footnotesize \@topsep30\p@\relax
  \advance\@topsep by -\baselineskip
  \item\relax
  \author@andify\authors
  \def\\{\protect\linebreak}%
  \textsc{\normalsize\textcolor{darkelectricblue}{\authors}}%
  \ifx\@empty\contribs
  \else
    ,\penalty-3 \space \@setcontribs
    \@closetoccontribs
  \fi
  \endtrivlist
  \endgroup
}
\def\@settitle{\begin{center}%
  \baselineskip14\p@\relax
    \bfseries
    \textsc{\Large\textcolor{charcoal}{\@title}}
  \end{center}%
}
\makeatother

\usepackage{enumitem}
\setlist[description]{%
  topsep=30pt,               
  itemsep=5pt,               
  font={\bfseries\sffamily\color{NavyBlue}}, 
}

\usepackage{fancyhdr}
\usepackage{lastpage}

\newcommand*\Title{\textcolor{bluepigment}{Improved Serre--Green--Naghdi equations}}
\newcommand*\Authors{\textcolor{bluepigment}{D.~Clamond, D.~Dutykh \& D.~Mitsotakis}}
\newcommand*{\plogo}{\textcolor{gray}{{\texttt{arXiv.org} / \textsc{hal}}}} 

\numberwithin{equation}{section}

\newtheorem{remark}{Remark}

\newcommand{\depth}{d}

\newcommand{\ub}{\breve{u}}
\newcommand{\um}{\bar{\/u}}

\newcommand{\ud}{\mathrm{d}}
\newcommand{\ui}{\mathrm{i}}
\newcommand{\ue}{\mathrm{e}}

\renewcommand{\O}{\mathcal{O}}

\renewcommand{\alpha}{\upalpha}

\newcommand{\phis}{\tilde{\phi}}

\renewcommand{\epsilon}{\varepsilon}

\newcommand{\vu}{\boldsymbol{u}}
\newcommand{\vx}{\boldsymbol{x}}
\newcommand{\nab}{\boldsymbol{\nabla}}

\newcommand{\cf}{\emph{cf.}\xspace}

\newcommand{\ie}{\emph{i.e.}\xspace}
\newcommand{\eg}{\emph{e.g.}\xspace}

\newcommand{\scal}{\boldsymbol{\cdot}}

\newcommand{\OD}[2]{\frac{\mathrm{D} #1}{\mathrm{D}\/#2}}

\newcommand{\half}{{\textstyle{1\over2}}}
\newcommand{\third}{{\textstyle{1\over3}}}
\newcommand{\sixth}{{\textstyle{1\over6}}}
\newcommand{\fourth}{{\textstyle{1\over4}}}
\newcommand{\twothird}{{\textstyle{2\over3}}}

\usepackage{acronym}
\acrodef{sgn}[SGN]{Serre--Green--Naghdi}
\acrodef{csgn}[cSGN]{classical Serre--Green--Naghdi}
\acrodef{msgn}[mSGN]{modified Serre--Green--Naghdi}

\begin{document}

\title[\Title]{Conservative modified Serre--Green--Naghdi equations with improved dispersion characteristics}

\author[D. Clamond]{Didier Clamond$^*$}
\address{Laboratoire J. A. Dieudonn\'e, Universit\'e de Nice -- Sophia Antipolis, Parc Valrose, 06108 Nice cedex 2, France}
\email{diderc@unice.fr}
\urladdr{http://math.unice.fr/~didierc/}
\thanks{$^*$ Corresponding author}

\author[D.~Dutykh]{Denys Dutykh}
\address{LAMA, UMR 5127 CNRS, Universit\'e Savoie Mont Blanc, Campus Scientifique, 73376 Le Bourget-du-Lac Cedex, France}
\email{Denys.Dutykh@univ-savoie.fr}
\urladdr{http://www.denys-dutykh.com/}

\author[D.~Mitsotakis]{Dimitrios Mitsotakis}
\address{Victoria University of Wellington, School of Mathematics, Statistics and Operations Research, PO Box 600, Wellington 6140, New Zealand}
\email{dmitsot@gmail.com}
\urladdr{http://dmitsot.googlepages.com/}


\begin{titlepage}
\thispagestyle{empty} 
\noindent
{\Large Didier \textsc{Clamond}}\\
{\it\textcolor{gray}{Universit\'e de Nice -- Sophia Antipolis, France}}\\[0.02\textheight]
{\Large Denys \textsc{Dutykh}}\\
{\it\textcolor{gray}{CNRS, Universit\'e Savoie Mont Blanc, France}}
\\[0.02\textheight]
{\Large Dimitrios \textsc{Mitsotakis}}\\
{\it\textcolor{gray}{Victoria University of Wellington, New Zealand}}\\[0.16\textheight]

\colorbox{Lightblue}{
  \parbox[t]{1.0\textwidth}{
    \centering\huge\sc
    \vspace*{0.7cm}
    
    \textcolor{bluepigment}{Conservative modified Serre$-$Green$-$Naghdi equations with improved dispersion characteristics}

    \vspace*{0.7cm}
  }
}

\vfill 

\raggedleft     
{\large \plogo} 
\end{titlepage}


\newpage
\thispagestyle{empty} 
\par\vspace*{\fill}   
\begin{flushright} 
{\textcolor{denimblue}{\textsc{Last modified:}} \today}
\end{flushright}


\newpage
\maketitle
\thispagestyle{empty}


\begin{abstract}

For surface gravity waves propagating in shallow water, we propose a variant of the fully nonlinear \textsc{Serre}--\textsc{Green}--\textsc{Naghdi} equations involving a free parameter that can be chosen to improve the dispersion properties. The novelty here consists in the fact that the new model conserves the energy, contrary to other modified \textsc{Serre}'s equations found in the literature. Numerical comparisons with the \textsc{Euler} equations show that the new model is substantially more accurate than the classical \textsc{Serre} equations, specially for long time simulations and for large amplitudes.

\bigskip
\noindent \textbf{\keywordsname:} Shallow water waves; improved dispersion; energy conservation. \\


\end{abstract}


\newpage
\tableofcontents
\thispagestyle{empty}


\clearpage
\section{Introduction}

Water waves in channels and oceans are usually described by the \textsc{Euler} equations. Due to their complexity, several approximate models have been derived in various wave regimes. Once a new mathematical model is proposed, the limits of its applicability have to be determined. In shallow water, the main restriction comes from the ratio between the characteristic wavelength $\lambda$ and the mean water depth $d$, the so-called shallowness parameter $\sigma\ =\ \depth/\lambda\ \ll\ 1$. Restrictions on the free surface elevation are characterised by the dimensionless parameter $\epsilon\ =\ a/d$, where $a$ is a typical amplitude ($\epsilon\sigma=a/\lambda$ is a wave steepness). Many approximate equations have been derived for waves in shallow water, such as the \textsc{Korteweg--de Vries} (KdV) equation \cite{KdV} for unidirectional waves, the \textsc{Saint-Venant} equations \cite{Wehausen1960} for bidirectional non-dispersive waves and many variants of the \textsc{Boussinesq} equations \cite{BCS, Boussinesq1877} for dispersive waves propagating in both directions. In addition to shallowness ($\sigma\ \ll\ 1$), KdV and \textsc{Boussinesq} equations assume small amplitudes (\eg $\epsilon\ =\ \O(\sigma^2))$ \cite{Johnson2004}.

Considering long waves propagating in shallow water but without assuming small amplitudes (\ie $\sigma\ \ll\ 1$ and $\epsilon\ =\ \O(1)$), \textsc{Serre} \cite{Serre1953} derived a so-called {\em fully-nonlinear weakly dispersive} system of equations  \cite{Wei1995, Wu2001a}  which, after further approximations, include the \textsc{Korteweg--de Vries}, \textsc{Saint-Venant} and \textsc{Boussinesq} equations as special cases. For steady flows, these equations were already known to \textsc{Rayleigh} \cite{LordRayleigh1876}. The \textsc{Serre} equations were independently rediscovered by Su and Gardner \cite{Su1969}, and again later by \textsc{Green}, \textsc{Laws} and \textsc{Naghdi} \cite{Green1974}. These nowadays popular equations constitute an asymptotic fully nonlinear model including all the terms up to order $\sigma^3$ into the momentum equation. \textsc{Serre}'s equations represent a substantial improvement with respect to the Boussinesq theory \cite{Castro-Orgaz2015}, but many shallow water phenomena involve significant dispersive effects that are not well described by \textsc{Serre}'s equations.

One possibility to improve the \textsc{Serre} model is by including the terms of order $\sigma^5$ (and higher-order terms) into the momentum equation. This program was accomplished for \textsc{Boussinesq} equations in, \eg, \cite{Madsen1991}. However, this modification makes the fifth-order (and higher-order) derivatives to appear in the model equations, making its numerical resolution (and thus its applicability) rather challenging. Actually, the numerical resolution of these high-order \textsc{Boussinesq}-like equations is slower (and less accurate) than that of the \textsc{Euler} equations, at least for simple (periodic) domains.

Another way of improving the classical model was coined by \textsc{Bona} and \textsc{Smith} \cite{BS} (see also \cite{Nwogu1993, BCS}). The idea consists in introducing a free parameter into the model that can be appropriately chosen to improve some of the desired properties. This can be achieved by replacing the depth-averaged velocity variable by the velocity of the fluid evaluated at at a certain depth in the bulk of fluid \cite{Madsen1998}. Most often, this parameter is chosen to optimise somehow the dispersion relation properties \cite{Madsen2002, Kim2009}. In general, the community is rather focused on various linear properties of the model, even if it is used later to simulate nonlinear waves.

Similar ideas have also been applied to the \textsc{Serre} equations with flat \cite{Dias2010, Dutykh2014c, Liu2005} and varying bottoms \cite{AntunesDoCarmo2013, Carmo2013, Castro-Orgaz2015}. Free parameters are obtained from  arbitrarily-weighted averages of different (but of same order) approximations of some quantities. However, these modified \textsc{Serre} equations invalidate one fundamental physical property: the conservation of the energy. Some variants also invalidate the \textsc{Galilean} invariance \cite{Duran2013}, that is a severe drawback for many applications. The same remarks apply as well to previous works on the improved \textsc{Boussinesq} equations \cite{Beji1996, Madsen1992}. Thus, one may improve the dispersive properties of the model but, on the other hand, loses the energy conservation property. For many applications, specially in the case of long time simulations, the disadvantages can be crucial, overriding all the possible advantages. In the present paper, we address this issue, proposing a method for deriving an improved version of the \textsc{Serre} equations that preserves the aforementioned nice properties of the original \textsc{Serre} model. For the sake of simplicity, the method is illustrated for 2D waves over a horizontal bottom, but generalisations to 3D and varying bottom can be obtained in an analogous manner.

The present paper is organised as follows. In Section~\ref{seccmSerre} a simple derivation of the classical \textsc{Serre} equations is presented. It is followed by the derivation of an already known one-parameter generalisation of these equations and its shortcomings are explained. In Section~\ref{seciSerre} we derive a new one-parameter generalisation of the \textsc{Serre} equations that conserve the energy. In Section~\ref{seccripar} we discuss several criteria for choosing the free parameter. Some numerical results are provided in Section \ref{secnum} demonstrating the advantages of the new \textsc{Serre}-like equations. Finally, the main conclusions, possible generalisations and perspectives are 
discussed in Section~\ref{secdisc}.


\section{Classical and modified Serre's equations}
\label{seccmSerre}

\begin{figure}
  \centering
  \includegraphics[width=0.75\textwidth,height=50mm]{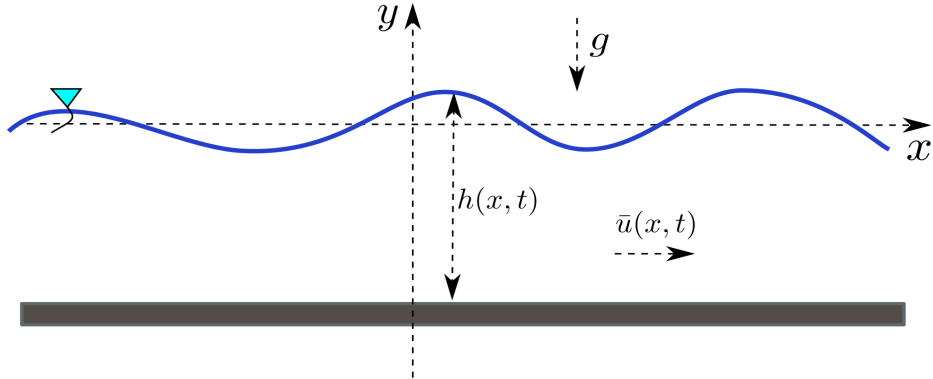}
  \caption{\small\em Sketch of the domain.}
  \label{fig:sketch}
\end{figure}

Here, we derive the classical \textsc{Serre} equations and a modified version of these equations with an additional free parameter using variational methods.


\subsection{Classical Serre's equations}
\label{ssecSerre}

In order to model irrotational two-dimensional long waves propagating in shallow water over horizontal bottom, one can approximate the velocity field by the ansatz  
\begin{equation}\label{defuvse}
  u(x,y,t)\ \approx\ \bar{u}(x,t), \qquad v(x,y,t)\ \approx\ -\,(y + d)\,\bar{u}_x,
\end{equation}
where $d$ is the water depth and $\bar{u}$ is the horizontal velocity averaged over the water column --- \ie, $\bar{u}\ =\ h^{-1}\int_{-d}^\eta u\,\ud y$, $h\ =\ \eta\ +\ d$ the total water depth --- $y\ =\ \eta$ and $y\ =\ 0$ being the equations of the free surface and of the still water level, respectively; a sketch of the fluid domain is depicted in the Figure~\ref{fig:sketch}. The horizontal velocity $u$ is thus (approximately) uniform along the water column and the vertical velocity $v$ is chosen so that the fluid incompressibility is valid. Note that the vorticity $\omega\ =\ v_x\ -\ u_y\ \approx\ -(y\ +\ d)\,\/\bar{u}_{xx}$ is not exactly zero, meaning that the potential flow is approximated by a rotational velocity field. The fact that the irrotationality is violated should not be more surprising than the violation of other relations, such as the isobarity of the free surface.

With the ansatz (\ref{defuvse}), the vertical acceleration is 
\begin{align*}
  \OD{v}{t}\ &=\ \frac{\partial\,v}{\partial\/t}\ +\ u\,\frac{\partial\,v}{\partial\/x}\ +\ v\,\frac{\partial\,v}{\partial\/y}\ \approx\  -\,v\,\bar{u}_x\ -\ (y + d)\,\OD{\bar{u}_x}{t}\ =\ \gamma\,\frac{y + d}{h}, 
\end{align*}
where $\gamma$ is the vertical acceleration at the free surface, \ie,
\begin{align}\label{defgamma}
  \gamma\ =\ \left.\OD{v}{t}\right|_{y=\eta}\ \approx\ h\left[\,\bar{u}_x^{\,2}\,-\,\bar{u}_{xt}\,-\,\bar{u}\,\bar{u}_{xx}\,\right]\,=\ 2\,h\,\bar{u}_x^{\,2}\ -\ h\,\partial_x\left[\,\bar{u}_{t}\,+\,\bar{u}\,\bar{u}_{x}\,\right]\,.
\end{align}
The kinetic and potential energies per water column, respectively $\mathscr{K}$ and $\mathscr{V}$, are similarly easily derived
\begin{equation*}
  \frac{\mathscr{K}}{\rho}\ =\ \int_{-d}^\eta\frac{u^2 + v^2}{2}\,\ud\/y\ \approx\ \frac{h\,\bar{u}^2}{2}\ +\ \frac{h^3\,\bar{u}_x^{\,2}}{6}, \qquad
  \frac{\mathscr{V}}{\rho}\ =\ \int_{-d}^\eta g\,(y + d)\,\ud\/y\ =\ \frac{g\,h^2}{2},
\end{equation*}
where $\rho\ >\ 0$ is the (constant) fluid density and $g\ >\ 0$ is the acceleration due to gravity (directed downward). A \textsc{Lagrangian} density $\mathscr{L}$ can then be introduced as the kinetic minus the potential energies plus a constraint for the mass conservation 
\begin{equation}\label{eq:dens}
  \frac{\mathscr{L}}{\rho}\ =\ \frac{h\,\bar{u}^2}{2}\ +\ \frac{h^3\,\bar{u}_x^{\,2}}{6}\ -\ 
\frac{g\,h^2}{2}\ +\,\left\{\,h_t\,+\left[\,h\,\bar{u}\,\right]_x\,\right\}\phis, 
\end{equation}
where $\phis$ is a \textsc{Lagrange} multiplier. Physically, $\phis$ is a (approximated) velocity potential expressed at the free surface.

The \textsc{Euler}--\textsc{Lagrange} equations for the action functional $\mathscr{S}\ =\ \iint\!\mathscr{L}\,\ud\/x\,\ud\/t$ 
are
\begin{align}
  \delta\phis:\quad & 0\ =\ h_t\ +\,\left[\,h\,\bar{u}\,\right]_x,\label{dLdphi}\\
  \delta\bar{u}:\quad & 0\ =\ \phis\,h_x\ -\ [\,h\,\phis\,]_x\ -\ \third\,[\,h^3\,\bar{u}_x\,]_x\ +\ h\,\bar{u}, \\
  \delta h:\quad & 0\ =\ \half\,\bar{u}^2\ -\ g\,h\ +\ \half\,h^2\,\bar{u}_x^{\,2}\ -\ \phis_t\ +\ \phis\,\bar{u}_x\ -\ [\,\bar{u}\,\phis\,]_x,
\end{align}
thence
\begin{align}
  \phis_x\ &=\ \bar{u}\ -\ \third\,h^{-1}\,[\,h^3\,\bar{u}_x\,]_x,\label{serphix}\\
  \phis_t\ &=\ \half\,h^2\,\bar{u}_x^{\,2}\ -\ \half\,\bar{u}^2\ -\ g\,h\ +\ \third\,\bar{u}\,h^{-1}\,[\,h^3\,\bar{u}_x\,]_x. \label{serphit}
\end{align}
Differentiating (\ref{serphit}) with respect to $x$ and using (\ref{serphix}) in order to eliminate $\phis$, after some elementary algebra, one obtains the surface tangential momentum equation (\cf Appendix~\ref{apptanmom} for the physical interpretation of this equation)
\begin{align}
  \partial_t\!\left[\,\bar{u}\,-\,\third\,h^{-1}(h^3\/\bar{u}_x)_x\,\right]\, +\ \partial_x\!\left[\,\half\,\bar{u}^2\,+\,g\,h\,-\,\half\,h^2\,\bar{u}_x^{\,2}\,-\,\third\,\bar{u}\,h^{-1}(h^3\/\bar{u}_x)_x\,\right]\, &=\ 0, \label{eqqdmbisse}
\end{align}
that can be written into the more familiar non-conservative form
\begin{equation}\label{eqqdmsemod}
  \bar{u}_t\ +\ \bar{u}\,\bar{u}_x\ +\ g\,h_x\ =\ -\,\third\,h^{-1}\,\partial_x\!\left[\,h^2\,\gamma\,\right]\,.
\end{equation}
After multiplication by $h$ and exploiting (\ref{dLdphi}), we derive the conservative equations for the momentum
\begin{align}
  \partial_t\!\left[\,h\,\bar{u}\,\right]\, +\ \partial_x\!\left[\,h\,\bar{u}^2\,+\,\half\,g\,h^2\,+\,\third\,h^2\,\gamma\,\right]\, &=\ 0, \label{eqqdmfluxse0}
\end{align}
that can be rewritten
\begin{align}
  \partial_t\!\left[\,h\,\bar{u}\,-\,\third(h^3\bar{u}_x)_x\,\right]\, +\ \partial_x\!\left[\,h\,\bar{u}^2\,+\,\half\,g\,h^2\,-\,\twothird\,h^3\,\bar{u}_x^{\,2}\,-\,\third\,h^3\,\bar{u}\/\bar{u}_{xx}\,-\,h^2\/h_x\/\bar{u}\/\bar{u}_x\,\right]\, &=\ 0. \label{eqqdmfluxsebis} 
\end{align}
From the two conservative equations (\ref{dLdphi}) and (\ref{eqqdmfluxse0}), an equation for the energy conservation can also be derived in the form
\begin{align}\label{eqenese}
  \partial_t\!\left[\,\half\,h\,\bar{u}^2\,+\,\sixth\,h^3\/\bar{u}_x^{\,2}\,+\,\half\,g\,h^2\,\right]\, +\ \partial_x\!\left[\,(\half\,\bar{u}^2\,+\,\sixth\,h^2\,\bar{u}_x^{\,2}+\,g\,h\,+\,\third\,h\,\gamma\,)\,h\,\bar{u}\,\right]\, &=\ 0.
\end{align}

In summary, we have just derived the following system of equations
\begin{align}
  h_t\ +\ \partial_x\!\left[\,h\,\bar{u}\,\right]\, &=\ 0, \label{eqmassse} \\
  \partial_t\!\left[\,h\,\bar{u}\,\right]\, +\ \partial_x\!\left[\,h\,\bar{u}^2\,+\,\half\,g\,h^2\,+\,\third\,h^2\,\gamma\,\right]\, &=\ 0, \label{eqqdmfluxse} 
\end{align}
together with (\ref{defgamma}), that are the \textsc{Serre} equations \cite{Serre1953} and are a special case of the \textsc{Green}--\textsc{Naghdi} equations \cite{Green1974}. Here, we refer to these equations as the \emph{classical \textsc{Serre}--\textsc{Green}--\textsc{Naghdi}} (cSGN) equations.

The derivation presented in this section is straightforward but, for readers more familiar with small parameter expansions, some additional material is given in Appendix~\ref{appasym}. Other variational derivations can be found, for instance, in \cite{Fedotova1996, Kim2001}.

Physically, equations (\ref{eqmassse}) and (\ref{eqqdmfluxse}) describe, respectively, the mass and momentum flux conservations. The physical interpretation of (\ref{eqqdmbisse}) have been debated in the literature. Actually, (\ref{eqqdmbisse}) is a conservative equation for the surface tangential momentum, as shown in Appendix~\ref{apptanmom} (see also \cite{Gavrilyuk2015} for an alternative derivation).


\subsection{Modified Serre's equations}
\label{ssecmSerre}

The cSGN equations describe long waves in shallow water, thus the horizontal and temporal derivatives are small quantities, \ie, $\partial_x\ \propto\ \O(\sigma)$ and $\partial_t\ \propto\ \O(\sigma)$, where $\sigma\ll1$ is of the order of the water depth divided by the characteristic wavelength (Appendix~\ref{appasym}). As a consequence, the left-hand side of (\ref{eqqdmsemod}) is of order one, while its right-hand side is of order three ($\gamma$ is of order two), while higher-order terms are neglected in this equation.

The definition (\ref{defgamma}) of $\gamma$ involving parts of the equation (\ref{eqqdmsemod}), the latter can be used to derive another approximation of the vertical acceleration at the free surface. Indeed, substituting the relation
\begin{equation*}
  \bar{u}_t\ +\ \bar{u}\,\bar{u}_x\ =\ -\,g\,h_x\ -\ \third\,h^{-1}\,\partial_x\!\left[\, h^2\,\gamma\,\right]\,,
\end{equation*}
into the definition of $\gamma$, one obtains to the same order of approximation (Appendix~\ref{appasym}) 
\begin{equation}\label{defaccvermod}
\gamma\ =\ 2\,h\,\bar{u}_x^{\,2}\ +\ g\,h\,h_{xx}\ +\ \O\!\left(\sigma^4\right),
\end{equation}
which is another expression for the vertical acceleration consistent with the order of approximation. It is however possible to obtain a more general system averaging the two expressions (\ref{defgamma}) and \eqref{defaccvermod} as 
\begin{equation}\label{defaccvermodgen}
  \gamma\ =\ 2\,h\,\bar{u}_x^{\,2}\ +\ \beta\,g\,h\,h_{xx}\ +\ (\beta-1)\,h\,\partial_x\!\left[\,\bar{u}_t\,+\,\bar{u}\,\bar{u}_x\,\right]\,+\ \O\!\left(\sigma^4\right),
\end{equation}
where $\beta$ is a parameter at our disposal. Thus, the modified \textsc{Serre} equations are
\begin{align}
  h_t\ +\ \partial_x\!\left[\,h\,\bar{u}\,\right]\, &=\ 0, \label{eqmasssem} \\
  \partial_t\!\left[\,h\,\bar{u}\,\right]\, +\ \partial_x\!\left[\,h\,\bar{u}^2\,+\,\half\,g\,h^2\,+\,\third\,h^2\,\gamma\,\right]\, &=\ 0, \label{eqqdmfluxsem} \\
  2\,h\,\bar{u}_x^{\,2}\ +\ \beta\,g\,h\,h_{xx}\ +\ (\beta-1)\,\,h\,\partial_x\!\left[\,\bar{u}_t\,+\,\bar{u}\,\bar{u}_x\,\right]\,&=\ \gamma, \label{eqveraccmod}
\end{align}
and the cSGN equations are recovered if $\beta\ =\ 0$. These modified SGN equations (mSGN), or similar ones, have been derived and studied before in the literature (see, \eg, \cite{AntunesDoCarmo2013, Dias2010, Liu2005}).The free parameter $\beta$ involved in the mSGN equation can be chosen such that it optimises the linear dispersion relation (see Section~\ref{seccripar}), but other criteria can be used to define $\beta$ \cite{Dutykh2014c}.

From the modified \textsc{Serre} equations, one can derive a secondary relation
\begin{align}\label{mserqdmflux}
\partial_t\!\left[\,h\,\bar{u}\,+\,\third\,(\beta-1)\,(h^3\bar{u}_x)_x\,\right]\, +\ 
\partial_x\!\left[\,h\,\bar{u}^2\,+\,\half\,g\,h^2\,+\,\third\,\beta\,g\,h^3\,h_{xx}\quad\right.\nonumber \\
\left.+\,\twothird\,(2\beta-1)\,h^3\,\bar{u}_x^{\,2}\,+\,\third\,(\beta-1)\,h^3\,\bar{u}\,\bar{u}_{xx}\,
+\,(\beta-1)\,h^2\,h_x\,\bar{u}\,\bar{u}_x\,\right]\, &=\ 0, 
\end{align}
that is the modified counterpart of (\ref{eqqdmfluxsebis}). However, equations analogous to (\ref{eqqdmbisse}) and (\ref{eqenese}) cannot be obtained if $\beta\ \neq\ 0$. This means, in particular, that the energy is not conserved for the mSGN equations and that a variational principle cannot be obtained (if $\beta\ \neq\ 0$). This is a serious drawback, specially for long time simulations and for theoretical investigations.

It should be noted that instead of using the generalised acceleration (\ref{defaccvermodgen}) into the momentum equation (\ref{eqqdmfluxse}), it could be used into the energy equation (\ref{eqenese}). This yields other modified \textsc{Serre}'s equations conserving the energy but not the momentum. Thus, the shortcomings of the mSGN are not addressed that way (if $\beta\ \neq\ 0$).

The mSGN equations (\ref{eqqdmfluxsem}) -- (\ref{mserqdmflux}) are asymptotically consistent with their cSGN counterparts for all values of the parameters $\beta$. However, the mSGN equations for the tangential momentum and for the energy are worse than asymptotically inconsistent: they are not conservative if $\beta\neq0$. Even worse, a variational principle for the mSGN equations cannot be obtained if $\beta\ \neq\ 0\,$. Thus, seeking for an improved shallow water model tuning the momentum equation, some physically crucial properties are lost, although the modifications are consistent from an asymptotic viewpoint.  

In the next section, we derive another set of modified \textsc{Serre}'s equations that have a free parameter and that conserve both the energy and the momenta. 


\section{Consistent modified Serre's equations}
\label{seciSerre}

In order to derive a better variant of the SGN equations, we modify the \textsc{Lagrangian} (\ref{eq:dens}) appropriately instead of modifying directly the cSGN equations, this modified \textsc{Lagrangian} being asymptotically consistent with (\ref{eq:dens}). Thus, the existence of a variational principle for the resulting approximate equations is automatically ensured, as well as conservative equations for the momentum, energy and tangential momentum at the surface.


\subsection{Modified Lagrangian}

The \textsc{Lagrangian} density (\ref{eq:dens}) for the cSGN equations can be rewritten replacing $h\bar{u}_x^{\,2}$ by its expression in terms of $\gamma$ according to \eqref{defgamma} such that: 
\begin{align*}
  \frac{\mathscr{L}}{\rho}\ =\ \frac{h\,\bar{u}^2}{2}\ +\ \frac{h^2\,\gamma}{12}\ +\ \frac{h^3}{12}\,\left[\,\bar{u}_{t}\,+\,\bar{u}\,\bar{u}_{x}\,\right]_x\,-\ \frac{g\,h^2}{2}\ +\,\left\{\,h_t\,+\left[\,h\,\bar{u}\,\right]_x\,\right\}\phis.
\end{align*}
Substituting the generalised expression \eqref{eqveraccmod} for $\gamma$, we obtain the new \textsc{Lagrangian} $\mathscr{L}_0^\prime$ as 
\begin{align}\label{Lbisori}
  \frac{\mathscr{L}_0^\prime}{\rho}\ &=\ \frac{h\,\bar{u}^2}{2}\ +\ \frac{h^3\,\bar{u}_x^{\,2}}{6}\ +\ \frac{\beta\,h^3}{12}\left[\,\bar{u}_{t}\,+\,\bar{u}\,\bar{u}_{x}\,+\,g\,h_x\,\right]_x\ -\ \frac{g\,h^2}{2}\ +\,\left\{\,h_t\,+\left[\,h\,\bar{u}\,\right]_x\,\right\}\phis \nonumber \\
  &=\ \frac{\mathscr{L}}{\rho}\ +\ \frac{\beta\,h^3}{12}\left[\,\bar{u}_{t}\,+\,\bar{u}\,\bar{u}_{x}\,+\,g\,h_x\,\right]_x.
\end{align}
Obviously, $\mathscr{L}$ and $\mathscr{L}_0^\prime$ differ by fourth-order terms --- the squared bracket in (\ref{Lbisori}) --- that are neglected in these approximations. These two \textsc{Lagrangians} have therefore the same order of approximation according to the asymptotic analysis (Appendix~\ref{appasym}). It should be noted that the squared bracket in (\ref{Lbisori}) has a simple generalisation in three dimensions, thus allowing a straightforward three-dimensional extension of this modified \textsc{Lagrangian} (Appendix~\ref{app3d}).

Integrating by parts the terms involving $\bar{u}_t$, $\bar{u}\bar{u}_x$ and $h_{xx}$, and using the mass conversation in order to remove the resulting term $h_t$, one gets
\begin{align*}
  h^3\,[\,\bar{u}_{t}\,+\,\bar{u}\,\bar{u}_{x}\,]_x\ &=\ \partial_t[\,h^3\,\bar{u}_x\,]\ -\ 3\,h^2\,h_t\,\bar{u}_x\ +\ h^3\,\bar{u}_x^{\,2}\ +\ h^3\,\bar{u}\,\bar{u}_{xx}\nonumber\\
  &=\ \partial_t[\,h^3\,\bar{u}_x\,]\ +\ \partial_x[\,h^3\,\bar{u}\,\bar{u}_x\,]\ +\ 3\,h^3\,\bar{u}_x^{\,2},\\
  h^3\,h_{xx}\ &=\ \partial_x[\,h^3\,h_x\,]\ -\ 3\,h^2\,h_x^{\,2}.
\end{align*}
Substituting these relations into (\ref{Lbisori}) and omitting the boundary terms $\partial_x[\cdots]$ and $\partial_t[\cdots]$ (since they do not contribute to the variational principle), we obtain the simplified modified \textsc{Lagrangian} 
\begin{align}\label{defLprime}
  \frac{\mathscr{L}^{\prime}}{\rho}\ =\ &\frac{h\,\bar{u}^2}{2}\ +\ \frac{(2+3\beta)\,h^3\,\bar{u}_x^{\,2}}{12}\ -\ \frac{g\,h^2}{2}\  -\ \frac{\beta\,g\,h^2\,h_{x}^{\,2}}{4}\ +\,\left\{\,h_t\,+\left[\,h\,\bar{u}\,\right]_x\,\right\}\phis.
\end{align}
With this \textsc{Lagrangian}, it follows that the modified kinetic and potential energy densities are, respectively,
\begin{equation*}
  \frac{\mathscr{K}^\prime}{\rho}\ =\ \frac{h\,\bar{u}^2}{2}\,+\,\frac{(2+3\beta)\,h^3\,\bar{u}_x^{\,2}}{12}, \qquad 
  \frac{\mathscr{V}^\prime}{\rho}\ =\ \frac{g\,h^2}{2}\left(\,1\,+\,\frac{\beta\,h_{x}^{\,2}}{2}\,\right), 
\end{equation*}
to be compared with the corresponding quantities in the cSGN (obtained when $\beta\ =\ 0$). In the cSGN, the kinetic energy is approximate (compared to the exact equations), while the potential energy is exact. On the other hand, both energies are approximate in (\ref{defLprime}) if $\beta\ \neq\ 0\,$. Having one (or more) exact relation in an approximated model is not necessarily a good thing; what really matters is to have a better approximation of the overall system of equations, as advocated in \cite{Clamond2009}. It is more important to have a better approximation of the complete system of equations than, \eg, an exact potential energy.


\subsection{Improved Serre's equations}

The \textsc{Euler}--\textsc{Lagrange} equations for the modified \textsc{Lagrangian} \eqref{defLprime} are
\begin{align}
  \delta\phis:\quad & 0\ =\ h_t\ +\,\left[\,h\,\bar{u}\,\right]_x,\label{dLGSWdphi} \\
  \delta\bar{u}:\quad & 0\ =\ h\,\bar{u}\ +\ \phis\,h_x\ -\,\left[\,h\,\phis\,\right]_x\ -\,\left(\third+\half\beta\right)\!\left[\,h^3\,\bar{u}_x\,\right]_x,\label{dLGSWdu} \\
  \delta h:\ & 0\ =\ \half\,\bar{u}^2\ -\ g\,h\ -\ \phis_t\ +\ \phis\,\bar{u}_x\ -\ [\,\bar{u}\,\phis\,]_x\nonumber\\
  &\qquad+\,\left(\half+\textstyle{3\over4}\beta\right)h^2\,\bar{u}_x^{\,2}\ -\ \half\,\beta\,g\,h\,h_{x}^{\,2}\,+\,\half\,\beta\,g\,[\,h^2\,h_x\,]_x,\label{dLGSWdh}
\end{align}
thence
\begin{align}
  \phis_x\ &=\ \bar{u}\ -\,\left(\third+\half\beta\right)h^{-1}\left[\,h^3\,\bar{u}_x\,\right]_x, \label{GSWphix} \\
  \phis_t\ &=\ \half\,\bar{u}^2\ -\ \bar{u}\,\phis_x\ -\ g\,h\ +\,\left(\half+\textstyle{3\over4}\beta\right)h^2\,\bar{u}_x^{\,2}\ +\ \half\,\beta\,g\,h\,[\,h\,h_x\,]_x. \label{GSWphit}
\end{align}
After differentiation of \eqref{GSWphit} with respect to $x$ and using \eqref{GSWphix}, we obtain the {\em improved \textsc{Serre}--\textsc{Green}--\textsc{Naghdi}} (iSGN) equations, written as
\begin{align}
  h_t\ +\ \partial_x\!\left[\,h\,\bar{u}\,\right]\, &=\ 0, \label{GSWmass}\\
  q_t\ +\ \partial_x\!\left[\,\bar{u}\,q\,-\,\half\,\bar{u}^2\,+\,g\,h-\left(\half+ \textstyle{3\over4}\beta\right)h^2\,\bar{u}_x^{\,2}\,-\,\half\,\beta\,g\,(\/h^2\/h_{xx}\/+\/h\/h_x^{\,2}\/)\,\right]\, &=\ 0, \label{GSWmomq} 
\end{align}
where $q\ =\ \phis_x$ is given by (\ref{GSWphix}). Equation (\ref{GSWmomq}) is the iSGN counterpart of the equation (\ref{eqqdmbisse}) for the cSGN, \ie (\ref{GSWmomq}) is a conservative equation for  the surface tangential momentum. Analogs to the cSGN equations can also be derived.

A non-conservative momentum equation is obtained directly from (\ref{GSWmomq})
\begin{align}\label{eqiSGNut}
  \bar{u}_t\ +\ \bar{u}\,\bar{u}_x\ +\ g\,h_x\ +\ \third\,h^{-1}\,\partial_x\!\left[\,h^2\,\Gamma\,\right]\, =\ 0,
\end{align}
where
\begin{equation*}
  \Gamma\ =\ \left(1+\textstyle{3\over2}\beta\right)h\left[\,\bar{u}_x^{\,2}\, -\,\bar{u}_{xt}\,-\,\bar{u}\,\bar{u}_{xx}\,\right]\,-\ \textstyle{3\over2}\,\beta\,g\left[\,h\,h_{xx}\,+\,\half\,h_x^{\,2}\,\right].
\end{equation*}
Notice that $\Gamma\ \neq\ \gamma$ if $\beta\ \neq\ 0\,$. After multiplication by $h$, the equation (\ref{eqiSGNut}) yields at once the conservation of the momentum flux
\begin{equation}\label{GSWqdmflux1}
  \partial_t\!\left[\,h\,\bar{u}\,\right]\, +\ \partial_x\!\left[\,h\,\bar{u}^2\, + \,\half\,g\,h^2\,+\,\third\,h^2\,\Gamma\,\right]\, =\ 0\,. 
\end{equation}
Finally, the requested equation for the energy conservation can also be derived in the form
\begin{align} 
  \partial_t\!\left[\,\half\,h\,\bar{u}^2\,+\,(\sixth+\fourth\beta)\,h^3\,\bar{u}_x^{\,2}\,+\,\half\,g\,h^2\left(\/1\/+\/\half\/\beta\/h_x^{\,2}\right)\right]&\nonumber  \\ 
  +\ \partial_x\!\left[\left\{\half\,\bar{u}^2\,+\,(\sixth+\fourth\beta)\,h^2\,\bar{u}_x^{\,2}\,+\,g\,h\left(\/1\/+\/\fourth\/\beta\/h_x^{\,2}\right)+\,\third\,h\,\Gamma\,\right\}h\,\bar{u}\,+\,\half\,\beta\,g\,h^3\,h_x\,\bar{u}_x\,\right]&\ =\ 0. \label{GSWene}
\end{align}

Thanks to the variational principle, we have derived a one-parameter generalisation of the \textsc{Serre} equations that retains important physical properties such as the momentum and energy conservations, as well as the \textsc{Galilean} invariance. These modified equations have a variational principle that is consistent with the original one according to the asymptotic analysis.


\subsection{Steady waves}
\label{sec:steady}

Since the iSGN equations are \textsc{Galilean} invariant, we can consider now steady solutions (\ie, independent of time in the frame of reference moving with the wave). For $2L$-periodic solutions, the mean water depth $d$ and the mean depth-averaged velocity $-c$ are
\begin{equation}\label{defdc}
  d\ =\,\left<\,h\,\right>\,=\ \frac{1}{2L}\int_{-L}^{L}h\,\ud\/x, \qquad 
  -\,c\,d\ =\,\left<\,h\,\bar{u}\,\right>\,=\ \frac{1}{2L}\int_{-L}^{L}h\,\bar{u}\,\ud\/x, 
\end{equation}
thus $c$ is the wave phase velocity observed in the frame of reference without mean flow.

With $h\ =\ h(x)$ and $\bar{u}\ =\ \bar{u}(x)\,$, the mass conservation (\ref{GSWmass}) straightforwardly yields
\begin{equation*}
  \bar{u}\ =\ -\,c\,d\left/\,h\right.,
\end{equation*}
and substitution into \eqref{GSWmomq} and \eqref{GSWqdmflux1} gives, respectively,  
\begin{align}
  \frac{(2+3\beta)\,\mathcal{F}\,(2\/h\/h_{xx}-h_{x}^{\,2})}{12\,(h/d)^2}\, -\, \frac{\beta\,(h\/h_{xx} + h_{x}^{\,2})}{2\,(d/h)}\,+\,\frac{\mathcal{F}\,d^2}{2\,h^2}\ +\ \frac{h}{d}\, &=\, 1\/+\/\frac{\mathcal{F}}{2}\/+\/\mathcal{C}_1, \label{eqqdmfluxseperm}\\
  \frac{(2+3\beta)\,\mathcal{F}\,(h\/h_{xx}-h_{x}^{\,2})}{3\,(h/d)}\, -\, \frac{\beta\,(2\/h\/h_{xx}+h_{x}^{\,2})}{2\,(d/h)^2}\,+\,\frac{2\,\mathcal{F}\,d}{h}\ +\ \frac{h^2}{d^2}\, &=\, 1\/+\/2\,\mathcal{F}\/+\/\mathcal{C}_2, \label{eqqdmbisseperm}
\end{align}
$\mathcal{F}\ =\ c^2/gd$ being a \textsc{Froude} number squared and where $\mathcal{C}_n$ ($n\ =\ 1,2$) are dimensionless integration constants to be determined from the relations (\ref{defdc}) ($\mathcal{C}_1\ =\ \mathcal{C}_2\ =\ 0$ for solitary waves). Eliminating $h_{xx}$ between (\ref{eqqdmfluxseperm}) and (\ref{eqqdmbisseperm}), after some straightforward algebra, one obtains the first-order ordinary differential equation
\begin{align}\label{odeper}
  \left(\frac{\ud\,h}{\ud\/x}\right)^{\!2}\ =\ \frac{\mathcal{F}\,-\,(1+\mathcal{C}_2 + 2\mathcal{F})\,(h/d)\,+\,(2+2\mathcal{C}_1+\mathcal{F})\,(h/d)^2\,-\,(h/d)^3}{\left(\third+\half\beta\right)\mathcal{F}\,-\,\half\,\beta\,(h/d)^3}.
\end{align}
The left-hand side being positive, so is the right-hand side. Its numerator and the denominator are therefore necessarily of the same sign.

The equation (\ref{odeper}) can be solved in parametric form with the help of \textsc{Jacobian} elliptic functions. However, we consider here only solitary waves, \ie, $h(\infty)\ =\ d$ and $\bar{u}(\infty)\ =\ -c$ thence $\mathcal{C}_1\ =\ \mathcal{C}_2\ =\ 0\,$. Thus, with $h(x)\ =\ d\ +\ \eta(x)\,$, we obtain the differential equation
\begin{align*}
  \left(\frac{\ud\,\eta}{\ud\/x}\right)^{\!2}\ =\ \frac{(\mathcal{F}-1)\,(\eta/d)^2\,-\,(\eta/d)^3}{\left(\third+\half\beta\right)\mathcal{F}\,-\,\half\,\beta\,(1+\eta/d)^3}.
\end{align*}
Introducing the change of independent variable $x\mapsto\xi$ such that
\begin{equation}\label{defxxi}
  x(\xi)\ =\ \int_0^\xi\left|\,\frac{(\beta+2/3)\mathcal{F}\,-\,\beta\,h^3(\xi')/d^3}{(\beta+2/3)\,\mathcal{F}\,-\,\beta}\,\right|^{1/2}\,\ud\/\xi',
\end{equation}
we obtain the classical equation for solitary waves
\begin{align*}
  \left(\frac{\ud\,\eta}{\ud\/\xi}\right)^{\!2}\ =\,\left|\frac{(\mathcal{F}-1)\,(\eta/d)^2\,-\,(\eta/d)^3}{\left(\third+\half\beta\right)\mathcal{F}\,-\,\half\,\beta}\right|,
\end{align*}
with a solution given in the parametric form
\begin{equation}\label{defyxi}
  \frac{\eta(\xi)}{d}\ =\ (\mathcal{F}-1)\,\operatorname{sech}^2\!\left(\frac{\kappa\,\xi}{2}\right), \qquad 
  (\kappa\/d)^2\ =\ \frac{6\,(\mathcal{F}-1)}{(2+3\beta)\,\mathcal{F}\,-\,3\,\beta},
\end{equation}
the wave amplitude being $a\ =\ \eta(0)\ =\ (\mathcal{F}\ -\ 1)\/d\,$. Substitution of (\ref{defyxi}) into (\ref{defxxi}) yields an expression for $x(\xi)$. This expression being very complicated, though an explicit closed-form, it is not given here since it is of little practical interest (numerical quadrature is more efficient).

It should be noted  that the cSGN and iSGN equations have the same relation between the phase velocity and the wave amplitude for steady solitary waves (\ie, the relation $c(a)$ is independent of $\beta$), but other relations are generally not independent of $\beta\,$.


\section{Criteria for choosing the free parameter}
\label{seccripar}

We present here some criteria for choosing the free parameter $\beta$ The possibilities listed below are by no mean the only ones that could be considered.


\subsection{Linear approximation}
\label{ssecapplin}

For infinitesimal waves, $\eta$ and $\bar{u}$ being both small, the iSGN equations yield the linear system of equations
\begin{align}
  \eta_t\ +\ d\,\bar{u}_x\ &=\ 0, \\
  \bar{u}_t\ -\,\left(\third+\half\beta\right)d^2\,\bar{u}_{xxt}\ +\ g\,\eta_x\ -\ \half\,\beta\,g\,d^2\,\eta_{xxx}\ &=\ 0.
\end{align}
Seeking for traveling waves of the form $\eta\ =\ a\cos k(x-ct)\,$, one obtains the (linear) dispersion relation
\begin{equation*}
  \frac{c^2}{g\,d}\ =\ \frac{2\,+\,\beta\,(k\/d)^2}{2\,+\,(\twothird+\beta)\,(k\/d)^2}\ \approx\ 1\ -\ \frac{(k\/d)^2}{3}\ +\,\left(\frac{1}{3}+\frac{\beta}{2}\right)\frac{(k\/d)^4}{3},
\end{equation*}
that should be compared with the exact relation
\begin{equation}\label{disrelexa}
  \frac{c^2}{g\,d}\ =\ \frac{\tanh(k\/d)}{k\/d}\ =\ 1\ -\ \frac{(k\/d)^2}{3}\ +\ \frac{2\,(k\/d)^4}{15}\ -\ \frac{17\,(k\/d)^6}{315}\ +\ \cdots.
\end{equation}
One can choose the parameter $\beta$ so that the linear dispersion relation is correct up to the highest possible order of its \textsc{Taylor} expansion around $k\ =\ 0\,$. Thus, with $\beta\ =\ \textstyle{2/15}\ \approx\ 0.1333$ the exact and iSGN dispersion relations coincide up to $k^4$ (but only up to $k^2$ if $\beta\ \neq\ 2/15$).

It should be noted that this choice improves the dispersion relation only in the vicinity of $k\/d\ =\ 0\,$, \ie, for very long infinitesimal waves. Another possibility is to minimise some norm of the difference between the exact and approximate linear dispersion relations over an interval $k\ \in\ [0,\, k_\text{max}]$ with $k_\text{max}$ chosen {\em a priori}.


\subsection{Decay of a solitary wave}
\label{ssecappsol}

In the far field, a solitary wave decays like $\ue^{-\kappa|x|}\,$, the trend parameter $\kappa\ >\ 0$ being related to the Froude number by the ``dispersion'' relation\footnote{Recall that we are considering the frame of reference moving with the wave. In a fixed frame, the wave travels with a constant phase speed $c$ and solitary waves decays, both in space and time, like $\ue^{-\kappa|x-ct|}\,$. Thus, relations (\ref{disrelsolser}) and (\ref{disrelsolexa}) are not dispersion relations for the evanescent modes, the latter decaying in space and being periodic in time.}
\begin{equation}\label{disrelsolser}
  \frac{c^2}{g\,d}\ =\ \frac{2\,-\,\beta\,(\kappa\/d)^2}{2\,-\,(\twothird+\beta)\,(\kappa\/d)^2}\ =\ 1\ +\ \frac{(\kappa\/d)^2}{3}\ +\,\left(\frac{1}{3}+\frac{\beta}{2}\right)\frac{(\kappa\/d)^4}{3}\ +\ \cdots,
\end{equation}
to be compared with the exact relation \cite{McCowan1891, Stokes1905}
\begin{equation}\label{disrelsolexa}
  \frac{c^2}{g\,d}\ =\ \frac{\tan(\kappa\/d)}{\kappa\/d}\ =\ 1\ +\ \frac{(\kappa\/d)^2}{3}\ +\ \frac{2\,(\kappa\/d)^4}{15}\ +\ \frac{17\,(\kappa\/d)^6}{315}\ +\ \cdots.
\end{equation}  
The main advantage of this approach compared to the previous one is that \eqref{disrelsolexa} is exact and valid for all amplitudes, while (\ref{disrelexa}) is exact only in the limit $a\ \to\ 0\,$. Note that (\ref{disrelsolexa}) can be simply deduced from (\ref{disrelexa}) by setting $k\ =\ \ui\kappa\,$.

The tangent function is meromorphic with single poles at $\kappa d\ =\ \pm\pi/2$, $\pm 3\pi/2, \cdots$. The relation \eqref{disrelsolser} has a single pole at $\kappa d\ =\ \pm\sqrt{6/(2+3\beta)}\,$. This pole is at $\kappa\/d\ =\ \pm\pi/2$ if 
\begin{equation*}
  \beta\ =\ \twothird\left(\,12\,\pi^{-2}\,-\,1\,\right)\ \approx\ 0.1439\,.
\end{equation*}
This choice of $\beta$ is close to the optimal value $2/15$ proposed above. The fact that both methods (\textsc{Taylor} expansion and meromorphic interpolation) produce similar optimal values for $\beta$ is an indication that $\beta=2/15$ is a good choice.


\subsection{Limiting wave}
\label{ssecapplim}

For exact steady solutions, we focus now on the vicinity of the crest of the highest waves. To this end, we consider the solitary wave solution given above without loss of generality since the analysis is local. \textsc{Taylor} expansions of (\ref{defxxi}) and (\ref{defyxi}) around the crest $\xi=0$ give
\begin{align*}
  x(\xi)\ &=\,\left[\frac{(2+3\beta)\,\mathcal{F}\,-\,3\,\beta\,\mathcal{F}^3}{(2+3\beta)\,\mathcal{F}\,-\,3\,\beta}\right]^{1\over2}\/\xi\ +\ \O(\xi^3), \qquad 
  \eta(\xi)\ =\ a\left[\,1\,-\,\fourth\,(\kappa\xi)^2\,\right]\,+\ \O(\xi^4).
\end{align*}
Peaked crests are found if $\ud x/\ud\xi\ =\ 0$ at $\xi\ =\ 0\,$. That happens if
\begin{equation*}
\mathcal{F}^2\ =\ 1\ +\ \twothird\,\beta^{-1}.
\end{equation*}
With this peculiar limiting \textsc{Froude} number, the slope at the crest is
\begin{equation*}
  \left.\frac{\ud\,\eta}{\ud\/x}\right|_{x=0^\pm}\,=\ \frac{\ud\,\eta}{\ud\/\xi}\left/\frac{\ud\,x}{\ud\/\xi}\right|_{\xi=0^\pm}\,=\ \mp\,\frac{\kappa\,a\,\sqrt{\,\mathcal{F}^2\,+\,\mathcal{F}\,+\,1}}{\sqrt{3}\,\mathcal{F}}\ =\ \mp\,\frac{a\,\sqrt{2\,/\,3\/\beta}}{a\,+\,d}.
\end{equation*}
In other words, if $\beta\ \neq\ 0\,$, we have a family of limiting waves parametrised by the parameter $\beta\,$. Various inner angles are obtained depending on the choice of $\beta\,$.

For $\beta\ =\ 2/15\,$, we have $\mathcal{F}\ =\ \sqrt{6}\approx2.45\,$, $a/d\ =\ \sqrt{6}\ -\ 1\ \approx\ 1.45$ and the crest has (approximately) a $28.4^\circ$ inner angle. This is therefore not a good approximation of the limiting solitary wave.

For $\beta\ =\ \twothird(12\pi^{-2} - 1)\,$, the crest inner angle is about $37.3^\circ\,$. This is thus a somehow better approximation but still far from the exact $120^\circ$ angle \cite{Stokes1880}.

From these peculiar choices, it can be seen that the iSGN system optimised for linear waves cannot describe equally well the highest wave. A $120^\circ$ inner angle is obtained if
\begin{equation*}
  \kappa\,a\,\sqrt{\,\mathcal{F}^2\,+\,\mathcal{F}\,+\,1}\ =\ \mathcal{F} 
  \qquad\text{or}\qquad a\,\sqrt{2\,/\,\beta}\ =\ a\,+\,d,
\end{equation*}
and in terms of $\beta\,$, this condition becomes
\begin{equation*}
  \frac{3\,\beta}{4}\ =\ \frac{\beta\,+\,\sqrt{\,2\,\beta\,}\,-\,1}{\beta\,-\,\sqrt{\,8\,\beta\,}\,+\,2}.
\end{equation*}
This relation gives $\beta\ \approx\ 0.34560$ thence $a/d\ \approx\ 0.711\,$, that is much closer to the exact value $a/d\ \approx\ 0.8332$ \cite{Maklakov2002}. However, with this value of $\beta\,$, the linear dispersion relation of infinitesimal waves is not at all improved. 

\begin{remark}
The iSGN model provides limiting values for wave heights, unlike the cSGN one, that is an interesting feature in some situations. However, this model is not designed for modelling extreme waves, choosing $\beta$ to improve the linear dispersion relation is probably the best choice for most simulations.
\end{remark}


\section{Numerical examples}
\label{secnum}

In this section, we present few numerical tests illustrating the performance of the iSGN model. The numerical methods used are briefly described in corresponding sections below. The aim here is just to show that the new model outperforms the classical one.

\begin{figure}
  \centering
  \vspace*{0.89em}
  \includegraphics[width=0.9\textwidth,height=10cm]{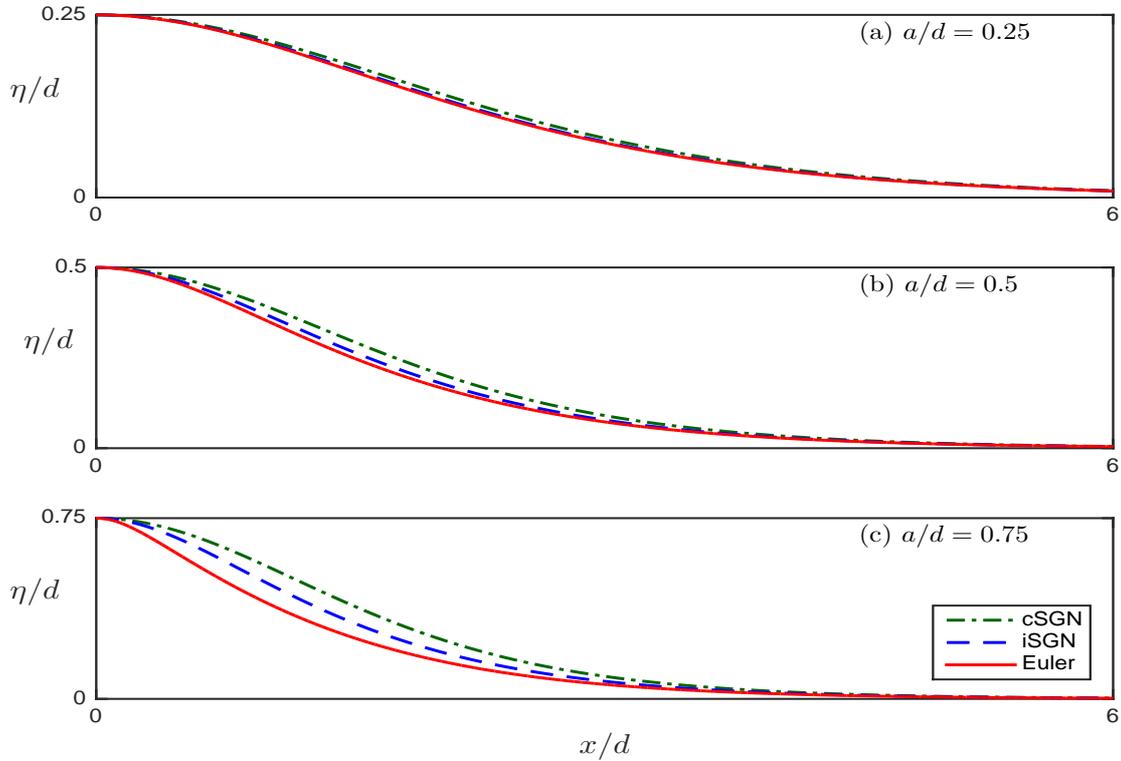}
  \caption{\small\em Comparison of steady solitary waves.} 
  \centerline{\scriptsize '---': \textsc{Euler}; '$--$': iSGN ($\beta=2/15$); '$-\cdot-$': cSGN ($\beta=0$).}
  \label{fig:0}
\end{figure}


\subsection{Steady solitary waves}

For the first test, we compare steady solitary wave solutions of the cSGN and the iSGN (with $\beta\ =\ 2/15$) models with the exact solution of the full \textsc{Euler} equations \cite{Clamond2012, Dutykh2013b}. The comparison of surface elevations for three different amplitudes is given in Figure~\ref{fig:0}. It can be seen that both models are good approximations for small amplitudes. For large amplitudes, however, the iSGN model is significantly better than the cSGN one. This result may be surprising because the value $\beta\ =\ 2/15$ is chosen to improve the dispersion of sinusoidal infinitesimal waves (see Section~\ref{ssecapplin}). This is not so surprising considering that this value of $\beta$ is close to the one optimising the the relation (\ref{disrelsolser}) relating the phase speed with the trend parameter (see Section~\ref{ssecappsol}).

\begin{figure}
  \centering
  \vspace*{0.80em}
  \includegraphics[width=0.9\textwidth,height=10cm]{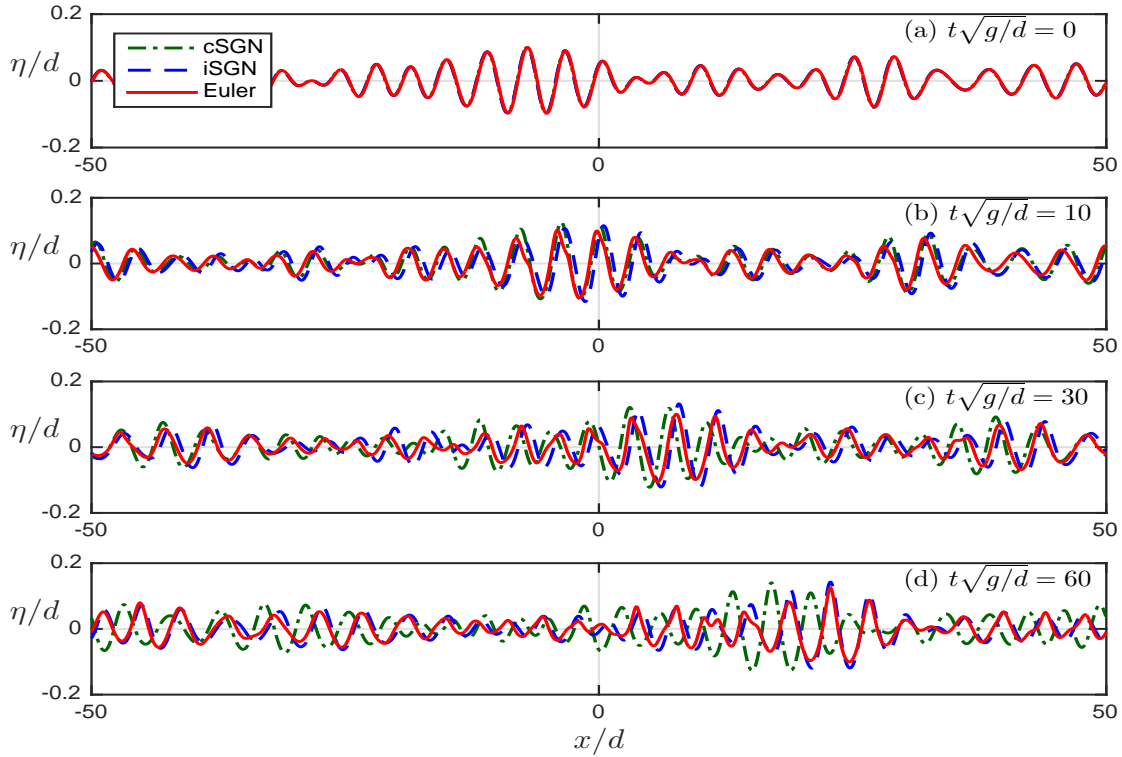}
  \caption{\small\em Evolution of a random initial condition in three different models.}
  \centerline{\scriptsize '---': \textsc{Euler}; '$--$': iSGN ($\beta=2/15$); '$-\cdot-$': cSGN ($\beta=0$).}
  \label{fig:1}
\end{figure}


\subsection{Random wave evolution}

We consider now a random initial condition (see upper Figure~\ref{fig:1}(a)) which is generated from the \textsc{Gaussian} \textsc{Fourier} power spectrum with random phases distributed uniformly in $]0,\, 2\pi]$ (see \cite{Dutykh2014c} for details where the same type of initial conditions were used). This initial wave is moderately steep ($a/d\ =\ 0.1$) but rather dispersive ($\sigma\ =\ 0.25$). Thus, this test focuses on the dispersive properties of the iSGN model.

In order to solve the iSGN equations on a periodic domain, we use the continuous \textsc{Galerkin}/Finite--Element method detailed in \cite{Mitsotakis2014}. The unsteady \textsc{Euler} equations are solved numerically using conformal mapping and pseudo-spectral discretisation \cite{Boyd2000,Li2004}.

\begin{figure}
  \centering
  \vspace*{0.89em}
  \includegraphics[width=0.9\textwidth,height=4cm]{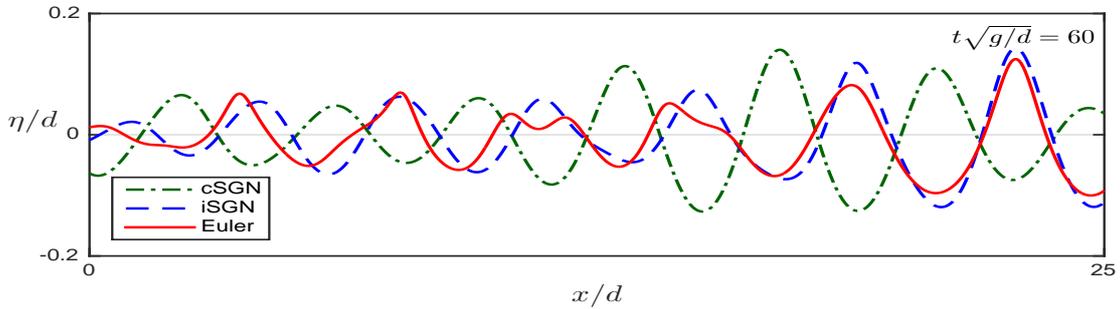}
  \caption{\small\em Magnification of Figure \ref{fig:1}(c).}
  \centerline{\scriptsize '---': \textsc{Euler}; '$--$': iSGN ($\beta=2/15$); '$-\cdot-$': cSGN ($\beta=0$).}
  \label{fig:2}
\end{figure}

The evolution of the generated random initial condition is shown in Figure~\ref{fig:1}. In Figure~\ref{fig:1}(\textit{a}) one can see that all three models are initialised with the same condition. During the evolution of this initial condition, the cSGN model starts lagging behind the reference solution given by the full \textsc{Euler} model. In Figure~\ref{fig:2}, we show a magnification of a portion of the computational domain at the final simulation time, where one can see that the iSGN equations commit a much lower phase error. It results a better prediction in the positions of the wave crests, that is important in applications focused on extreme waves.

\section{Discussion}
\label{secdisc}

In this paper, we presented a novel model for fully nonlinear long waves in shallow waters. These equations generalise the \textsc{Serre} equations in the sense that the new model contains a free parameter. The value of this parameter can be used to optimise, for instance, the linear dispersion relation properties. However, it is possible to use this extra degree of freedom to optimise the linear shoaling \cite{Beji1996} or any other desired characteristics.

In all the modified shallow water models we are aware of, the introduction of an extra parameter leads to the violation of the energy conservation. For some models, the \textsc{Galilean} invariance is violated too. Here, we succeeded in deriving a new model that preserves these properties, thanks to the use of a variational formalism based on \textsc{Hamilton} principle \cite{Clamond2009}. Instead of tweaking the system of PDEs, we modified the corresponding \textsc{Lagrangian} functional. This approach allows to preserve the underlying variational structure as well. Moreover, we showed that this model repeats another feature of the full \textsc{Euler} equations: the existence of limiting waves that are not included in the \textsc{Serre} equations.

We performed some unsteady simulations that clearly show the importance of the iSGN equations compared to the classical ones. The fact that the energy is conserved in the iSGN model is very important for long-time simulations. Also, we have shown that the improved dispersion relation leads to a better description of large solitary waves.

The main goal of this study was to make a proof of principle. However, the developments and ideas presented above can be generalised to non-flat bottom and in three dimensions (see Appendix~\ref{app3d}), for example. The derivations and examples presented here are sufficient to demonstrate the advantages of our approach.

Tweaking the vertical acceleration at the surface, we easily obtained a one-parameter generalisation of the classical \textsc{Serre} equation. Modifying other quantities, in the spirit of \cite{Dias2010}, one may obtain a multi-parameter generalisation of these equations. These extra free parameters can be tuned to improve even further the model equations. These generalisations are also left to future investigations.

We conclude this discussion by noting that, in addition to the \textsc{Galilean} invariance and the energy conservation, the dispersion-improved iSGN equations have the same order of derivatives and are thus not more difficult to solve numerically (similar algorithms and running times) compared to the classical \textsc{Serre} equations, unlike the high-order \textsc{Boussinesq}-like equations. This is an interesting feature for practical applications.


\appendix
\section{Asymptotic derivation}
\label{appasym}

For the horizontal velocity $u$, the most general solution of the \textsc{Laplace} equation satisfying the seabed impermeability (\cite{Lagrange1781}) is 
\begin{equation}\label{eq:lagr}
  u\ =\ \cos[(y + d)\partial_x]\,\ub\ =\ \ub\ -\ \half\,(y + d)^2\,\ub_{xx}\ +\ \textstyle{1\over24}\,(y+d)^4\,\ub_{xxxx}\ -\ \cdots,
\end{equation}
where $\ub(x,t)\ =\ u(x,y\!=\!-d,t)$ is the velocity at the bottom. Assuming long waves in shallow water means that $\partial_x\ =\ \O(\sigma)$ and, these waves having finite phase velocities, $\partial_t\ =\ \O(\sigma)$. Thence, the relation defining the depth-averaged horizontal velocity $\um$, \ie 
\begin{equation*}
  \um\ =\ \ub\ -\ \sixth\,h^2\,\ub_{xx}\ +\ \textstyle{1\over5!}\,h^4\,\ub_{xxxx}\ +\ \O\!\left(\sigma^6\right),
\end{equation*}
can be solved for $\ub$ as
\begin{equation*}
  \ub\ =\ \um\ +\ \sixth\,h^2\,\um_{xx}\ +\ \textstyle{7\over360}\,h^4\,\um_{xxxx}\ +\ \textstyle{1\over9}\,h^3\,h_x\,\um_{xxx}\ +\ \textstyle{1\over18}\,h^2\,(hh_x)_x\,\um_{xx}\ +\ \O\!\left(\sigma^6\right).
\end{equation*}
Thus, the horizontal velocity \eqref{eq:lagr} can be represented equivalently to $\O(\sigma^4)$ as
\begin{equation*}
  u\ =\ \um\ +\ \sixth\left[\,h^2\,-\,3\,(y + d)^2\,\right]\um_{xx}\ +\ \O\!\left(\sigma^4\right).
\end{equation*}
Similarly, fulfilling the fluid incompressibility and the bottom impermeability, the vertical velocity $v$ is 
\begin{equation*}
  v\ =\ -\,(y + d)\,\um_x\ -\ \sixth\,(y + d)\left[\,h^2\,-\,(y + d)^2\,\right]\um_{xxx}\ -\ \third\,(y+d)\,h\,h_x\,\um_{xx}\ +\ \O\!\left(\sigma^5\right),
\end{equation*}
thence at the free surface $\tilde{v}\ =\ -h\um_x\ -\ \third h^2 h_x \um_{xx}\ +\ \O(\sigma^5)\,$.

Assuming finite velocity and surface elevation, we take $\um = \O(\sigma^0)$ --- thus $\phis\ =\ \O(\sigma^{-1})$ because $u\ \approx\ \phis_x$) and $\eta\ =\ \O(\sigma^0)$ --- and the depth-integrated kinetic and potential energies are
\begin{equation*}
  \frac{\mathscr{K}}{\rho}\ =\, \int_{-d}^\eta\frac{u^2+v^2}{2}\,\ud\/y\ =\ \frac{h\,\bar{u}^2}{2}\ +\ \frac{h^3\,\bar{u}_x^{\,2}}{6}\ +\ \O\!\left(\sigma^4\right), \qquad
  \frac{\mathscr{V}}{\rho}\ =\,\int_{-d}^\eta g\,(y + d)\,\ud\/y\ =\ \frac{g\,h^2}{2}. 
\end{equation*}
The incompressibility and the bottom impermeability being identically satisfied, the \textsc{Hamilton} principle can be reduced to the \textsc{Lagrangian} density
\begin{align*}
  \frac{\mathscr{L}}{\rho}\ =\ \frac{h\,\bar{u}^2}{2}\ +\ \frac{h^3\,\bar{u}_x^{\,2}}{6}\ -\ \frac{g\,h^2}{2}\ +\ \left\{\,h_t\,+\,\left[\,h\um\,\right]_x\,\right\}\phis\ +\ \O\!\left(\sigma^4\right),
\end{align*}
which, after neglecting higher-order terms, corresponds exactly to the \textsc{Lagrangian} density in the action integral (\ref{eq:dens}).

It should be noted that instead of enforcing the mass conservation, one could enforce the impermeability of the free surface. These two approaches are equivalent here, however. Indeed, the incompressibility and the bottom impermeability are identically fulfilled with (\ref{defuvse}), fulfilling the surface impermeability yields the mass conservation, and vice versa.


\section{Tangential momentum at the free surface}
\label{apptanmom}

Let $\phi$ be the velocity potential of an irrotational motion, i.e., $u=\phi_x$, $v=\phi_y$. 
Denoting with `tildes' the quantities written at the free surface\footnote{Note that, e.g., 
$\tilde{u}=\widetilde{\,\phi_x\/}\neq\phis_x=\tilde{u}+\tilde{v}\eta_x$.} and exploiting the 
surface impermeability, we have the relation 
\begin{equation}
\widetilde{\,\phi_t\,}\ =\ \phis_t\ -\ \eta_t\,\tilde{v}\ =\ \phis_t\ -\ \tilde{v}^2\ 
+\ \eta_x\,\tilde{u}\,\tilde{v}\ =\ \phis_t\ -\ \tilde{v}^2\ +\,\left(\phis_x-\tilde{u}\right)
\tilde{u}.
\end{equation}
The Bernoulli equation at the free surface can then be rewritten
\begin{align}
\phis_t\ +\,\left(\phis_x-\tilde{u}\right)\tilde{u}\ +\ g\,\eta\ +\ \half\,\tilde{u}^2\ 
-\ \half\,\tilde{v}^2\ =\ 0.
\end{align}
After differentiation with respect of $x$, this equation becomes
\begin{equation}
\partial_t\!\left[\,\phis_x\,\right]\, +\ \partial_x\!\left[\,g\,\eta\, +\, \half\,\tilde{u}^2\, 
-\,\half\,\tilde{v}^2\,+\left(\phis_x - \tilde{u}\right)\tilde{u}\,\right]\,=\ 0.
\end{equation}
Equation (\ref{eqqdmbisse}) is recovered using the ansatz \eqref{defuvse} and the relation (\ref{serphix}). 
Therefore, Equation (\ref{eqqdmbisse}) represents the horizontal derivative of the Bernoulli equation at the 
free surface, i.e., the surface tangential component of the momentum (Euler) equation for irrotational flow. 
Alternative derivations of this result can be found in \cite{Gavrilyuk2015}.


\section{Generalisation in three dimensions}
\label{app3d}

The generalisation of our approach in three dimensions is straightforward. Let $\vx\ =\ (x_1,\,x_2)$ the horizontal \textsc{Cartesian} coordinates and $\vu\ =\ (u_1,\,\,u_2)$ the horizontal velocity field. A shallow water ansatz fulfilling the fluid incompressibility and the (horizontal) bottom impermeability is
\begin{equation}\label{defuvse3d}
  \vu(\vx,y,t)\ \approx\ \bar{\vu}(\vx,t), \qquad 
  v(\vx,y,t)\ \approx\ -\,(y+d)\,\nab\scal\bar{\vu},
\end{equation}
where $\nab$ is the horizontal gradient. From this ansatz, one can derive the `irrotational' \textsc{Green}--\textsc{Naghdi} equations \cite{Clamond2012, Kim2001}
\begin{align*}
  h_t\ +\ \nab\scal\left[\,h\,\bar{\vu}\,\right]\,&=\ 0, \\
  \bar{\vu}_t\ +\ \nab\left[\frac{|\bar{\vu}|^2}{2}\right]\, +\ g\,\nab h\ +\ \frac{\nab\scal\left[\,h^2\,\gamma\,\right]}{3\,h}\ &=\ \frac{\bar{\vu}\scal\nab h}{3}\,\nab\!\left[\,h\,\nab\scal\bar{\vu}\,\right]\,-\,\left[\,\bar{\vu}\scal\nab\left(\frac{h\nab\scal\bar{\vu}}{3}\right)\right]\nab h, 
\end{align*}
where $\gamma\ =\ h\{(\nab\scal\bar{\vu})^2\ -\ \nab\scal\bar{\vu}_t\ -\ \bar{\vu}\scal\nab[\nab\scal\bar{\vu}]\}$ is the vertical acceleration at the free surface.

These equations can be straightforwardly derived from the \textsc{Lagrangian} density
\begin{align*}
  \frac{\mathscr{L}}{\rho}\ =\  \frac{h\,|\bar{\vu}|^2}{2}\ +\ \frac{h^3\,(\nab\scal\bar{\vu})^2}{6}\ -\ \frac{g\,h^2}{2}\ +\,\left\{\,h_t\,+\,\nab\scal\left[\,h\,\bar{\vu}\,\right]\,\right\}\phis.
\end{align*}
Along the lines of the two-dimensional case (see equation \eqref{Lbisori}), an obvious alternative \textsc{Lagrangian} is
\begin{align*}
  \mathscr{L}'\ =\ \mathscr{L}\ +\ \textstyle{1\over12}\,\rho\,\beta\,h^3\,\nab\scal\left[\,\bar{\vu}_t\,+\,\half\,\nab|\bar{\vu}|^2\,+\,g\,\nab h\,\right],
\end{align*} 
where $\beta$ is a dimensionless constant at our disposal. According to the asymptotic analysis of Appendix~\ref{appasym}, $\mathscr{L}$ is obtained neglecting terms of order $\sigma^4$ (and higher) and we have 
\begin{align*}
  \nab\scal\left[\,\bar{\vu}_t\,+\,\half\,\nab|\bar{\vu}|^2\,+\,g\,\nab h\,\right]\,=\ 0\ +\ \O(\sigma^4).
\end{align*}
Thus, clearly, $\mathscr{L}$ and $\mathscr{L}'$ differ by fourth-order terms, so they are both consistent to the same order of approximation considered here. From $\mathscr{L}'$, one can easily derive modified equations of motion with a free parameter that can be chosen to improve the linear dispersion relation, for example. These equations yield automatically the energy conservation. The derivations are left to the reader.
 
There are several 3D generalisations of the \textsc{Serre} equations that have been debated in the literature \cite{DellIsola2011}. We picked one of them in order to illustrate that the improvement method can be easily used in three dimensions; our goal here is not to explore in details all the possible 3D cases. It is also clear that the approach presented here can be easily used to improve models in presence of a varying bottom and surface tension, for example.


\subsection*{Acknowledgments}
\addcontentsline{toc}{subsection}{Acknowledgments}

D.~\textsc{Dutykh} \& D.~\textsc{Clamond} acknowledge the support from CNRS under the PEPS InPhyNiTi project \textsc{FARA}. D.~\textsc{Mitsotakis} was supported by the \textsc{Marsden Fund} administered by the Royal Society of New Zealand.


\addcontentsline{toc}{section}{References}
\bibliographystyle{abbrv}
\bibliography{biblio}

\begin{thebibliography}{10}

\bibitem{AntunesDoCarmo2013}
J.~S. {Antunes Do Carmo}.
\newblock {Boussinesq and Serre type models with improved linear dispersion
  characteristics: Applications}.
\newblock {\em J. Hydr. Res.}, 51(6):719--727, dec 2013.

\bibitem{Carmo2013}
J.~S. {Antunes Do Carmo}.
\newblock {Extended Serre Equations for Applications in Intermediate Water
  Depths}.
\newblock {\em The Open Ocean Engineering Journal}, 6(1):16--25, aug 2013.

\bibitem{Beji1996}
S.~Beji and K.~Nadaoka.
\newblock {A formal derivation and numerical modelling of the improved
  Boussinesq equations for varying depth}.
\newblock {\em Ocean Engineering}, 23(8):691--704, nov 1996.

\bibitem{BCS}
J.~L. Bona, M.~Chen, and J.-C. Saut.
\newblock {Boussinesq equations and other systems for small-amplitude long
  waves in nonlinear dispersive media. I: Derivation and linear theory}.
\newblock {\em J. Nonlinear Sci.}, 12:283--318, 2002.

\bibitem{BS}
J.~L. Bona and R.~Smith.
\newblock {A model for the two-way propagation of water waves in a channel}.
\newblock {\em Math. Proc. Camb. Phil. Soc.}, 79:167--182, 1976.

\bibitem{Boussinesq1877}
J.~V. Boussinesq.
\newblock {Essai sur la th{\'{e}}orie des eaux courantes}.
\newblock {\em M{\'{e}}moires pr{\'{e}}sent{\'{e}}s par divers savants {\`{a}}
  l'Acad. des Sci. Inst. Nat. France}, XXIII:1--680, 1877.

\bibitem{Boyd2000}
J.~P. Boyd.
\newblock {\em {Chebyshev and Fourier Spectral Methods}}.
\newblock New York, 2nd edition, 2000.

\bibitem{Castro-Orgaz2015}
O.~Castro-Orgaz and W.~H. Hager.
\newblock {Boussinesq- and Serre-type models with improved linear dispersion
  characteristics: applications}.
\newblock {\em J. Hydraulic Res.}, 53(2):282--284, mar 2015.

\bibitem{Clamond2012}
D.~Clamond and D.~Dutykh.
\newblock
  http://www.mathworks.com/matlabcentral/fileexchange/39189-solitary-water-wave,
  2012.

\bibitem{Clamond2009}
D.~Clamond and D.~Dutykh.
\newblock {Practical use of variational principles for modeling water waves}.
\newblock {\em Phys. D}, 241(1):25--36, 2012.

\bibitem{DellIsola2011}
F.~Dell'Isola and S.~Gavrilyuk.
\newblock {\em {Variational Models and Methods in Solid and Fluid Mechanics}},
  volume 535 of {\em CISM International Centre for Mechanical Sciences}.
\newblock Springer Vienna, Vienna, 2011.

\bibitem{Dias2010}
F.~Dias and P.~Milewski.
\newblock {On the fully-nonlinear shallow-water generalized Serre equations}.
\newblock {\em Phys. Lett. A}, 374(8):1049--1053, 2010.

\bibitem{Duran2013}
A.~Duran, D.~Dutykh, and D.~Mitsotakis.
\newblock {On the Galilean Invariance of Some Nonlinear Dispersive Wave
  Equations}.
\newblock {\em Stud. Appl. Math.}, 131(4):359--388, nov 2013.

\bibitem{Dutykh2013b}
D.~Dutykh and D.~Clamond.
\newblock {Efficient computation of steady solitary gravity waves}.
\newblock {\em Wave Motion}, 51(1):86--99, jan 2014.

\bibitem{Dutykh2014c}
D.~Dutykh, D.~Clamond, and D.~Mitsotakis.
\newblock {Adaptive modeling of shallow fully nonlinear gravity waves}.
\newblock {\em RIMS K{\^{o}}ky{\^{u}}roku}, 1947(4):45--65, 2015.

\bibitem{Fedotova1996}
Z.~I. Fedotova and E.~D. Karepova.
\newblock {Variational principle for approximate models of wave hydrodynamics}.
\newblock {\em Russ. J. Numer. Anal. Math. Modelling}, 11(3):183--204, 1996.

\bibitem{Gavrilyuk2015}
S.~Gavrilyuk, H.~Kalisch, and Z.~Khorsand.
\newblock {A kinematic conservation law in free surface flow}.
\newblock {\em Nonlinearity}, 28(6):1805--1821, jun 2015.

\bibitem{Green1974}
A.~E. Green, N.~Laws, and P.~M. Naghdi.
\newblock {On the theory of water waves}.
\newblock {\em Proc. R. Soc. Lond. A}, 338:43--55, 1974.

\bibitem{Johnson2004}
R.~S. Johnson.
\newblock {\em {A Modern Introduction to the Mathematical Theory of Water
  Waves}}.
\newblock Cambridge University Press, 2004.

\bibitem{Kim2009}
G.~Kim, C.~Lee, and K.-D. Suh.
\newblock {Extended Boussinesq equations for rapidly varying topography}.
\newblock {\em Ocean Engineering}, 36(11):842--851, aug 2009.

\bibitem{Kim2001}
J.~W. Kim, K.~J. Bai, R.~C. Ertekin, and W.~C. Webster.
\newblock {A derivation of the Green-Naghdi equations for irrotational flows}.
\newblock {\em J. Eng. Math.}, 40(1):17--42, 2001.

\bibitem{KdV}
D.~J. Korteweg and G.~de~Vries.
\newblock {On the change of form of long waves advancing in a rectangular
  canal, and on a new type of long stationary waves}.
\newblock {\em Phil. Mag.}, 39(5):422--443, 1895.

\bibitem{Lagrange1781}
J.-L. Lagrange.
\newblock {M{\'{e}}moire sur la th{\'{e}}orie du mouvement des fluides}.
\newblock {\em Nouv. M{\'{e}}m. Acad. Berlin}, 196, 1781.

\bibitem{Li2004}
Y.~A. Li, J.~M. Hyman, and W.~Choi.
\newblock {A Numerical Study of the Exact Evolution Equations for Surface Waves
  in Water of Finite Depth}.
\newblock {\em Stud. Appl. Maths.}, 113:303--324, 2004.

\bibitem{Liu2005}
Z.~B. Liu and Z.~C. Sun.
\newblock {Two sets of higher-order Boussinesq-type equations for water waves}.
\newblock {\em Ocean Engineering}, 32(11-12):1296--1310, aug 2005.

\bibitem{LordRayleigh1876}
J.~W.~S. {Lord Rayleigh}.
\newblock {On Waves}.
\newblock {\em Phil. Mag.}, 1:257--279, 1876.

\bibitem{Madsen2002}
P.~A. Madsen, H.~B. Bingham, and H.~Liu.
\newblock {A new Boussinesq method for fully nonlinear waves from shallow to
  deep water}.
\newblock {\em J. Fluid Mech.}, 462:1--30, 2002.

\bibitem{Madsen1991}
P.~A. Madsen, R.~Murray, and O.~R. Sorensen.
\newblock {A new form of the Boussinesq equations with improved linear
  dispersion characteristics}.
\newblock {\em Coastal Engineering}, 15:371--388, 1991.

\bibitem{Madsen1998}
P.~A. Madsen and H.~A. Schaffer.
\newblock {Higher-Order Boussinesq-Type Equations for Surface Gravity Waves:
  Derivation and Analysis}.
\newblock {\em Phil. Trans. R. Soc. Lond. A}, 356:3123--3184, 1998.

\bibitem{Madsen1992}
P.~A. Madsen and O.~R. Sorensen.
\newblock {A new form of the Boussinesq equations with improved linear
  dispersion characteristics. Part 2. A slowly-varying bathymetry}.
\newblock {\em Coastal Engineering}, 18:183--204, 1992.

\bibitem{Maklakov2002}
D.~Maklakov.
\newblock {Almost-highest gravity waves on water of finite depth}.
\newblock {\em European Journal of Applied Mathematics}, 13:67--93, 2002.

\bibitem{McCowan1891}
J.~McCowan.
\newblock {On the solitary wave}.
\newblock {\em Phil. Mag. S.}, 32(194):45--58, 1891.

\bibitem{Mitsotakis2014}
D.~Mitsotakis, B.~Ilan, and D.~Dutykh.
\newblock {On the Galerkin/Finite-Element Method for the Serre Equations}.
\newblock {\em J. Sci. Comput.}, 61(1):166--195, feb 2014.

\bibitem{Nwogu1993}
O.~Nwogu.
\newblock {Alternative form of Boussinesq equations for nearshore wave
  propagation}.
\newblock {\em J. Waterway, Port, Coastal and Ocean Engineering}, 119:618--638,
  1993.

\bibitem{Serre1953}
F.~Serre.
\newblock {Contribution {\`{a}} l'{\'{e}}tude des {\'{e}}coulements permanents
  et variables dans les canaux}.
\newblock {\em La Houille blanche}, 8:374--388, 1953.

\bibitem{Stokes1880}
G.~G. Stokes.
\newblock {Supplement to a paper on the theory of oscillatory waves}.
\newblock {\em Mathematical and Physical Papers}, 1:314--326, 1880.

\bibitem{Stokes1905}
G.~G. Stokes.
\newblock {\em {The outskirts of the solitary wave}}, volume~5.
\newblock University Press, Cambridge, 1905.

\bibitem{Su1969}
C.~H. Su and C.~S. Gardner.
\newblock {KdV equation and generalizations. Part III. Derivation of the
  Korteweg-de Vries equation and Burgers equation}.
\newblock {\em J. Math. Phys.}, 10:536--539, 1969.

\bibitem{Wehausen1960}
J.~V. Wehausen and E.~V. Laitone.
\newblock {Surface waves}.
\newblock {\em Handbuch der Physik}, 9:446--778, 1960.

\bibitem{Wei1995}
G.~Wei, J.~T. Kirby, S.~T. Grilli, and R.~Subramanya.
\newblock {A fully nonlinear Boussinesq model for surface waves. Part 1. Highly
  nonlinear unsteady waves}.
\newblock {\em J. Fluid Mech.}, 294:71--92, 1995.

\bibitem{Wu2001a}
T.~Y. Wu.
\newblock {A unified theory for modeling water waves}.
\newblock {\em Adv. App. Mech.}, 37:1--88, 2001.

\end{thebibliography}

\end{document}